\documentstyle[12pt,psfig]{article}
\def\simlt{\stackrel{<}{{}_\sim}}
\def\simgt{\stackrel{>}{{}_\sim}}
\begin{document}

\title{The Higgs Boson Mass as a Probe of the Minimal Supersymmetric
Standard Model}

\author{M. Carena$^1$, P.H. Chankowski$^2$,\\
S. Pokorski$^{2,3}$, C.E.M. Wagner$^3$
\\
\\
$^1$ Fermi National Accelerator Laboratory, \\
P.O. Box 500, Batavia, IL 60510, USA
\\
$^2$ Institute of Theoretical Physics, Warsaw University\\
ul. Ho\.za 69, 00--681 Warsaw, Poland
\\
$^3$ CERN, TH Division, CH-1211 Geneva 23, Switzerland.
}

\maketitle

\vspace{-12cm}
\begin{flushright}
{\bf FERMILAB-PUB--98/146-T}\\
{\bf IFT--98/23}\\
{\bf CERN-TH/98-148}\\
\end{flushright}
\vspace{12cm}

\begin{abstract}
Recently, the LEP collaborations have reported a lower bound on a
Standard Model-like Higgs boson of order 89 GeV. We discuss 
the implications of this bound  for the minimal supersymmetric
extension of the Standard Model (MSSM). In particular, we show that the
lower bound on $\tan\beta$, which can be obtained from the
presently allowed
Higgs boson mass value, becomes stronger than the one set by
the requirement of perturbative consistency of the theory
up to scales of order $M_{GUT}$ (associated with the infrared fixed-point 
solution of the top quark Yukawa coupling) in a large fraction of
the allowed parameter space. The potentiality of future LEP2 searches
to further probe
the MSSM parameter space is also discussed.
\end{abstract}

\newpage

One of the most striking predictions of the minimal supersymmetric extension
of the Standard Model (the MSSM) is the existence of a light, 
${\cal O}(100$~GeV), Higgs particle.
The supersymmetric prediction for the range of the lighter CP-even
Higgs boson mass is nicely consistent with the fits to the electroweak
precision data (for recent fits see \cite{ERPI,PRECDATA}). However, the
existence
of the Higgs boson has not yet been directly established experimentally and
the search for it continues to be the main goal of LEP2. The absence of such
a light Higgs particle would eventually rule out  low energy supersymmetry in
its minimal version.
The present experimental lower bound for its mass enters into the region
most relevant for the MSSM. It is, therefore, quite timely to discuss the
constraints on the MSSM derived from the present and
near-future expected lower
bounds on the Higgs boson mass $M_h$ or by the potential discovery
of a light Higgs boson with  a mass $M_h$.
One of the most interesting aspects of this question
is the lower bound on the parameter $\tan\beta$
($\tan\beta=v_2/v_1$, where $v_1$ and $v_2$ are the two Higgs boson doublet
vacuum expectation values), which can be derived within this context.
Considering the MSSM as a
low-energy effective theory, the
bounds on $\tan\beta$ depend on the physical stop masses
(and their mixing angle) but do not depend on any theoretical assumption,
e.g. on the pattern of soft terms at the GUT scale, or, more generally,
on the actual mechanism that communicates supersymmetry breaking to the
observable sector.

Supersymmetric extensions of the Standard Model provide a framework for a
consistent link between low-energy physics and physics at the GUT scale.
Since the top Yukawa coupling is not asymptotically free, the requirement
of perturbative consistency of the theory up to the scale $M_{GUT}$ puts
a strong and very interesting bound on the top-quark Yukawa
coupling at the scale of the top-quark mass, $h_t(m_t)$ \cite{IRFIX}. The
bound depends slightly on the mass spectrum of the MSSM and can be somewhat
altered by the
presence of extra matter, e.g. ${\bf 5} + \bar{\bf 5}$ vector-like
multiplets at some intermediate scale $M_I$.
With the measured value $m^{pole}_t=173.9\pm5.2$ GeV~\cite{TOPMASS}, and by
using the relation
\begin{eqnarray}
m^{pole}_t={1\over\sqrt2}h_t(m_t){v \sin\beta} + ... 
\label{eqn:mtyukrel}
\end{eqnarray}
(where $v^2\equiv v^2_1+v^2_2=4M^2_W/(g^{\prime2}+g^2) + ...$, and the
ellipses stand for perturbative corrections), the upper
bound on $h_t(m_t)$ can also be translated into a lower bound on $\tan\beta$.

In this letter we compare the bounds on $\tan\beta$ obtained for a given
value of $M_h$ within the low-energy MSSM with the bound on $\tan\beta$ derived
from the requirement of perturbative consistency of the theory up to the
scale $M_{GUT}$. We shall show that even the present experimental limit on
$M_h$ implies a bound on $\tan\beta$, which is well above the perturbativity
bound for a large range of stop masses (and mixings). The infrared fixed-point
scenario, associated with the values of the
top-quark Yukawa coupling  close to the
perturbative upper bound remains consistent with the present limit on
$M_h$ only for large values of the heavier stop mass, large stop mass
splitting and large mixing angle.

Stronger lower bounds on $M_h$ imply more stringent lower
bounds on $\tan\beta$, which are
consistent with the infrared fixed-point scenario only for heavier and heavier
stops; eventually the two bounds no longer intersect each other. This is
consistent with the well-known upper bound on $M_h$ obtained in the infrared
fixed-point scenario in the minimal supergravity model, with universal
soft SUSY-breaking terms at the GUT scale.
As shown in ref. \cite{CAOLPOWA} (and
recently confirmed in a further study~\cite{ESHA}), in this case
$M_h\simlt98$ GeV for $M_{\tilde t_i}\simlt1$ TeV,
where $M_{\tilde t_i}$ are the physical stop masses.
Our general analysis shows
that in the unconstrained low-energy MSSM the above limit can only be slightly
relaxed, by at most a few  GeV.
For instance, $M_h\simeq103$ GeV is consistent
with the infrared fixed point of $h_t$ for stop masses of the order of 1~TeV, 
but
only for very large and positive values of the stop mixing angle
and/or a top mass close to its upper 1$\sigma$ range.

In general,  taking into account the full structure of the stop mass
matrix, the lighter CP-even 
Higgs boson mass in the MSSM is parametrized by
\begin{eqnarray}
M_h = M_h
\left(M_A,\tan\beta,m_t,M_{\tilde t_1},M_{\tilde t_2},A_t,\mu,...\right)
\end{eqnarray}
where $A_t$ and $\mu$ determine the mixing angle of the stops
(as well as some of their trilinear couplings to the Higgs bosons) and the
ellipses stand for other parameters whose effects are not dominant
(e.g. the gaugino mass parameters, or the sbottom sector parameters, which
become relevant only for large values of $\tan\beta > 10$).

The maximal $M_h$ is always obtained for $M_A\gg M_Z$ (in
practice, the bound is saturated for  $M_A\simgt250$ GeV).
In this limit one gets from the effective potential
approach (see ref. \cite{HABER} for details) a particularly simple
result for the one-loop corrected $M_h$ \cite{HE}:
\begin{eqnarray}
M^2_h=M^2_Z\cos^22\beta+{3\alpha\over4\pi s^2_W}{m^4_t\over M^2_W}
\left[\log\left({M^2_{\tilde t_2}M^2_{\tilde t_1}\over m^4_t}\right)
+\left({M^2_{\tilde t_2}-M^2_{\tilde t_1}\over 4 m^2_t}\sin^22\theta_{\tilde t}
\right)^2\right.\nonumber
\end{eqnarray}
\begin{eqnarray}
\times\left. f(M^2_{\tilde t_2},M^2_{\tilde t_1})
+ {M^2_{\tilde t_2}-M^2_{\tilde t_1}\over2m^2_t}\sin^22\theta_{\tilde t}
\log\left({M^2_{\tilde t_2}\over M^2_{\tilde t_1}}\right)\right]
\label{eqn:mhha}
\end{eqnarray}
where $f(x,y)=2 - (x + y)/(x - y)\log(x/y)$. For large $M_A$, the
dependence on the parameters $A_t$ and $\mu$ always appears in the
combination $\tilde{A}_t = A_t - \mu/\tan\beta$,~\footnote{The quoted
formula stays valid even in the case that $A_t$ and/or $\mu$
develop complex phases, provided $\tilde{A}_t$
is replaced everywhere 
by $|\tilde{A}_t|\equiv|A_t-\mu^*/\tan\beta|$.}
through the
dependence on the left--right
stop mixing angle $\theta_{\tilde t}$,
\begin{equation}
\sin2\theta_{\tilde t}={2m_t\tilde{A}_t\over M^2_{\tilde t_2} -
M^2_{\tilde t_1}} .
\label{sin2t}
\end{equation}
The two-loop corrections to
$M_h$ are typically ${\cal O}(20$\%) of the one-loop
corrections and are negative~\cite{HEHO}--\cite{HHH}.
They are taken into account in our numerical
results (we use the method proposed in \cite{CAESQUWA}).
Eq. (\ref{eqn:mhha}) is, however, very useful  for
the qualitative understanding of the final
results.  As can be seen from Eq. (\ref{eqn:mhha}),
for a given lower limit on $M_h$, larger values of the radiative
corrections to $M_h$ are required for lower values of $\tan\beta$ (i.e.
for smaller values of the tree level part of $M_h^2$).
Thus, maximizing the radiative corrections to $M_h$ sets a
lower bound on $\tan\beta$ as a function of $M_{\tilde t_1}$,
$M_{\tilde t_2}$ and $\sin2\theta_{\tilde t}$.
It is hence interesting to analyse the dependence 
of Eq. (\ref{eqn:mhha}) on $\sin2\theta_{\tilde t}$.

For fixed $M_{\tilde t_1}$, $M_{\tilde t_2}$, the mass of the
lighter $CP$-even Higgs boson is a quadratic function of
$\sin^2 2 \theta_{\tilde t}$,
\begin{equation}
M_h^2 = F(M_{\tilde t_i}) + G(M_{\tilde t_i}) \sin^2 2 \theta_{\tilde t}
+ H(M_{\tilde t_i}) \sin^4 2\theta_{\tilde t},
\end{equation}
with $G(M_{\tilde t_i}) > 0$ and $H(M_{\tilde t_i}) < 0$. Hence, there
is a maximum value of $M_h^2$, which is obtained for
\begin{eqnarray}
\left(\sin^2 2\theta_{\tilde t}\right)^{\rm max} ={-4 \; m^2_t
\log\left(M^2_{\tilde t_2}/M^2_{\tilde t_1}\right)
\over2\left(M^2_{\tilde t_2}-M^2_{\tilde t_1}\right) -
\left(M^2_{\tilde t_2} + M^2_{\tilde t_1}\right)
\log\left(M^2_{\tilde t_2}/M^2_{\tilde t_1}\right)}.
\label{sin2tmax}
\end{eqnarray}
For small values of the stop mass splitting, the value of
$\sin^2 2\theta_{\tilde t}$ that is obtained from the above expression
is larger than 1. Indeed, when the difference
between the stop masses is much smaller than their sum,
\begin{equation}
\left(\sin^2 2\theta_{\tilde t}\right)^{max}\simeq
\frac{12 m_t^2\left(M^2_{\tilde t_2} + M^2_{\tilde t_1}\right)}
{\left(M^2_{\tilde t_2} - M^2_{\tilde t_1}\right)^2}.
\end{equation}
In these cases the physical value of the Higgs boson mass
takes
its maximal value for $\sin^2 2\theta_{\tilde t}=1$, 
that is for equal values
of the diagonal entries $M^2_{\tilde t_R}$,
$M^2_{\tilde t_L}$, of the stop square mass matrix.
Moreover, in such cases,
zero mixing angle represents a minimum of the Higgs boson mass value.

For $\sin^2 2\theta_{\tilde t}=1$ the stop masses are given by
$M^2_{\tilde t_{1,2}}=M_{\rm SUSY}^2\mp m_t\tilde{A}_t$, where 
$M_{SUSY}^2 = M^2_{\tilde t_L} = M^2_{\tilde t_R}$.
In this case, it is possible to get a simple expression 
for $M_h^2$ including
the dominant two-loop leading-logarithmic corrections, written in
terms of $\tilde{A}_t$ and $M_{SUSY}$. In the limit
$(m_t \tilde{A}_t)/M_{\rm SUSY}^2\ll1$, it is given by \cite{CAESQUWA}
\begin{eqnarray}
M_h^2&\simeq& M_Z^2\cos^2 2\beta\left(1-{3\over8\pi^2}
\frac{m_t^2}{v^2} L_t\right)\nonumber \\
&+&{3\over4\pi^2}{m_t^4\over v^2}\left[{1\over2}\tilde{X}_t + L_t
+{1\over16\pi^2}\left({3\over2}\frac{m_t^2}{v^2}-8g^2_3
\right)\left(\tilde{X}_t L_t+L_t^2\right)\right],
\label{mhsm}
\end{eqnarray}
where $L_t\equiv\log\left(M_{\rm SUSY}^2/m_t^2\right)$,
\begin{equation}
\label{escala}
\tilde X_t={2\tilde{A}^2_t\over M_{\rm SUSY}^2}
\left(1-{\tilde{A}_t^2\over12M_{\rm SUSY}^2}\right)
\end{equation}
and $m_t$ and $g^2_3\equiv4\pi\alpha_3$ are the running top mass and
the strong gauge coupling evaluated at the scale $m_t$, respectively.
The expression (\ref{mhsm})
for $M_h^2$ has a maximum at $\tilde{A}_t=\sqrt6M_{\rm SUSY}$. Thus, for
moderate values of the stop mass splitting, the Higgs boson mass is
maximized by keeping equal values of the diagonal entries in the stop mass
matrix ($M^2_{\tilde t_R}\approx M^2_{\tilde t_L}$)
and is approximately given by Eq. (\ref{mhsm}).

For sufficiently large values of the stop mass splitting, the value of
$(\sin2\theta_{\tilde t})^{max}$ given by Eq. (\ref{sin2tmax}) becomes
lower than 1. Even in these cases, the maximal Higgs boson mass is obtained
for rather large values of $\tilde{A}_t$. To see this we can take, for
example, the case in which the heavier stop mass is of order 1 TeV and
the lighter stop mass is of the order of the top quark mass.  From
Eq. (\ref{sin2tmax}) we get
\begin{equation}
\left(\sin^2 2\theta_{\tilde t} \right)^{max} \simeq
10{m_t^2\over M^2_{\tilde t_2}},
\end{equation}
corresponding to a value of $|\sin2\theta_{\tilde t}|\simeq 0.6$.
Comparing the above expression with Eq. (\ref{sin2t}), we get
\begin{equation}
|\tilde{A}_t|\simeq1.5 M_{\tilde t_2}.
\end{equation}
Hence, as stated above, large values of $\tilde{A}_t$ are necessary 
to maximize $M_h^2$, even
in the case of very large splitting of the stop masses. 
From the value of $\sin2\theta_{\tilde t}$
it is also clear that in this case
the splitting in the left- and right-handed stop
masses is crucial for generating the difference in physical masses of the
heavier and lighter stops.

The computation of the Higgs boson
mass is still affected by theoretical uncertainties, most notably,
those associated with the two-loop finite threshold corrections to the
effective quartic couplings of the Higgs potential. Recently, partial
diagrammatic two-loop computation of the Higgs mass 
has been performed~\cite{HEHOWE}. 
Taking the appropriate limit, the values
obtained by this method are in agreement with our results 
within a range of 2-3 GeV. 
We take these differences as the estimates of the uncertainty
of the computed $M_h$. In order to take this uncertainty into account
and to remain on the conservative side, in all cases discussed below 
we have  lowered the bound 
by $2$ GeV with respect to  the actually
considered Higgs boson mass limit. 
We have also considered low values
of the chargino and neutralino masses 
(of the order of 200~GeV) to minimize
their negative effects on the Higgs masses.

Our numerical results are shown in Figs.~1--3.
In Fig. 1, we plot the lower bounds
on $\tan\beta$, following from the present experimental limit of 
89.5 GeV~\cite{HLIMIT} on a Standard Model (SM)-like Higgs boson mass
($M_A\simgt250$ GeV)~\footnote{The Higgs boson mass bound quoted here has
been obtained from a combined analysis of three of the four LEP experiments.
The individual ALEPH analysis, not included in the combination, leads
to a bound of order 87.9 GeV. The combination of the Higgs search data
of the four experiments would lead to a slightly larger value of the
Higgs mass bound, of order $M_Z$~\cite{JANOT}.}, as a
function of $M_{\tilde t_2}$ (the heavier stop mass) for several values of
the stop mass splitting $\Delta M_{\tilde t}\equiv M_{\tilde t_2}-
M_{\tilde t_1}$. For a given $M_{\tilde t_2}$ and $\Delta M_{\tilde t}$,
a scan over $\sin\theta_{\tilde t}$ is performed in order to find the lowest
value of $\tan\beta$ allowed by the limit imposed on $M_h$. For values
of the stop mass splitting of order 400~GeV or larger, the minimal value
of $\tan\beta$ is obtained for $|\sin 2 \theta_{\tilde t}| < 1$ and,
therefore, the left- and
right-handed stop mass parameters $M_{\tilde t_L}$, $M_{\tilde t_R}$
begin to differ, but the value of
$\tilde{A}_t$ always remains
larger than $M_{\tilde t_2}$, in agreement with our
discussion above. We have also verified that the
values of $\tilde{A}_t$ that maximize the Higgs boson mass, after the
dominant leading logarithm two-loop corrections to the effective
potential are included, are in good agreement with the ones
obtained from the one-loop expression,
Eqs. (\ref{sin2t}) and (\ref{sin2tmax}).

In the same figure we also show the bounds on $\tan\beta$, obtained from
the requirement of perturbative consistency of the theory up to the grand
unification scale~\footnote{The dependence
of the lower bound on $\tan\beta$ on the precise value of $M_{GUT}$ and
$\alpha_s(M_Z)$ (we use $\alpha_s(M_Z)=0.118$) is not significant for our
purpose.} $M_{GUT} \simeq 2\times10^{16}$ GeV. In the MSSM, for sufficiently
large values of the top-quark Yukawa coupling at the GUT scale,
its low-energy values
are governed by the quasi-infrared fixed-point solution \cite{IRFIX}
\begin{equation}
\left(h_t^2(Q)\right)_{IR} \simeq{8\over9}g_3^2(Q)~~~~~~~
{\rm for} ~Q\sim {\cal O}(100 {\rm GeV}).
\label{YFP}
\end{equation}
In order to obtain the physical top-quark  mass,
we compute the RG evolution of the Yukawa coupling
from $M_{GUT}$ down to the scale $Q = m_t$.
The physical top-quark mass is then calculated by
including all finite corrections in Eq.~(\ref{eqn:mtyukrel}).
The SM part of the corrections to  Eq.~(\ref{eqn:mtyukrel})
is dominated by the gluon contribution and, 
at the scale $Q=m_t$, is known up
to  ${\cal O}(\alpha_s^2)$~\cite{GRBRGRSCH}.
The corresponding SM one-loop
corrections to Eq.~(\ref{eqn:mtyukrel}) proportional to the
top-quark Yukawa coupling
are small, of the order of
the two-loop QCD ones\cite{HEKN}. The
one-loop supersymmetric particle corrections to Eq.~(\ref{eqn:mtyukrel})
have been calculated in~\cite{BAMAPI}, 
and their relevance for the correct definition
of the infrared fixed point  
solution has been stressed in Refs.~\cite{Nir,ESHA}.
For values of the heavier stop mass and/or gluino masses much larger
than the top-quark mass, they are dominated by two terms: the first contains
large logarithmic factors, 
which can as well be taken into
account by introducing appropriate step functions in the RGEs \cite{CAPIRA}:
\begin{equation}
{d\over dt}h_t^2\simeq{h_t^2\over(4\pi)^2}
\left\{\left[{9\over2}+\theta_{\tilde{Q}}+{1\over2}\theta_{\tilde{U}}
\right] h_t^2
-\left[8-{4\over3}\left(\theta_{\tilde{Q}}+\theta_{\tilde{U}}\right)
\theta_{\tilde{g}}\right]g_3^2\right\}
\label{htrun}
\end{equation}
\begin{equation}
{d\over dt}g_3^2 = {g_3^4\over(4\pi)^2}
\left\{-7 + 2\theta_{\tilde{g}} + {1\over2}
\left[2\theta_{\tilde{Q}}+\theta_{\tilde{U}}+\theta_{\tilde{D}}\right]\right\}
\label{g3run},
\end{equation}
where $t=\log(Q^2/M_Z^2)$, $\theta_{\tilde{X}}=\theta(Q-M_{\tilde{X}})$ and
$\tilde Q$, $\tilde U$ and $\tilde D$ stand for the left-handed doublet,
right-handed up and right-handed down squarks respectively. For simplicity, we
have  assumed
that the squark masses are generation independent. When
the logarithmic factors are large, they must be resummed, which we have
done in our computations~\cite{CH2}.
The second dominant term contains the non-logarithmic effects,
\begin{equation}
\frac{\delta m_t}{m_t} \simeq
-\frac{2 \alpha_3}{3 \pi} \tilde{A}_t m_{\tilde{g}}
\times I(m_{\tilde g}^2,M_{\tilde t_1}^2,M_{\tilde t_2}^2) ,
\label{finite}
\end{equation}
where
\begin{equation}
I(a,b,c) = \frac{ a \; b \; \log(a/b) + a \; c \; \log(c/a) +
b \; c \; \log(b/c)}
{(a-b) \;(b-c) \;(a - c)} \simeq  {\cal O}\left( \frac{1}
{{\rm max}(a,b,c)} \right)\nonumber.
\end{equation}

{ }From the above expressions,
the dependence of $\tan\beta$ obtained at the
infrared fixed point solution on the sparticle spectrum may be qualitatively
understood.
For instance, if all supersymmetric particle masses take equal values
$M_{\rm SUSY}\gg m_t$, the running of the top quark Yukawa coupling from the
scale $M_{\rm SUSY}$ to the scale $Q\sim{\cal O}(m_t)$ will be governed by
Eqs. (\ref{htrun}), (\ref{g3run}) with all $\theta_{\tilde{X}}=0$. In this case
the top quark Yukawa coupling becomes smaller at high energies compared to
the case in which sparticles are light (all $\theta_{\tilde{X}}=1$ from
$Q\sim{\cal O}(m_t)$ up to $Q=M_{GUT}$),
because in the former case the  coefficients
in Eq.~(\ref{htrun}) between $Q = m_t$ and $Q = M_{\rm SUSY}$ cause
a slower increase of $h_t$. This implies a  
smaller lower bound on $\tan\beta$ for heavier sparticles.
The mass of the gluino also has important effects on the bounds:
if it is much lower than the stop masses, it makes the strong gauge
coupling less asymptotically free in the low-energy effective theory,
slowing the evolution of the top Yukawa coupling to large values with
respect to the case $m_{\tilde g}\approx M_{\tilde t_i}$, implying 
again a
smaller lower bound on $\tan\beta$. On the other hand, the gluino
mass also controls the non-leading logarithmic 
corrections, Eq.~(\ref{finite}),
which become
larger for heavier gluinos and larger values of the stop mixing 
parameters and, in certain regions of parameter space, can be of the order 
of or larger than the leading-logarithmic  
corrections. In Fig. 1 we plot the maximal
and minimal values of $\tan\beta$ at the fixed-point solution, which
are obtained for a heavy gluino mass, of the order of the heavier stop one,
and for $\sin 2 \theta_{\tilde t} = \pm 1$, and also for heavy
and light gluinos when the stop mass splitting vanishes.

The comparison of the lower bounds on $\tan\beta$
obtained from the lower limits on  $M_h$
and from perturbativity of the top Yukawa coupling is striking.
In order to interpret the results correctly one should take
into account that, for
the same stop mass splitting, the bounds on $\tan\beta$ coming
from the limits on the Higgs boson mass
move rapidly up when the mixing 
parameter $|\tilde{A}_t|$ is lowered, while the bound set by the 
perturbativity of the top Yukawa coupling
moves  down (up) for positive (negative) values
of $\tilde{A}_t$ (by definition $m_{\tilde g} >0$).
Already with the present lower limit on the Higgs boson mass, and
for $M_{\tilde t_2} \simlt 1$ TeV, the  bounds on $\tan\beta$ coming from
the experimental limit on $M_h$ 
dominate over those coming from the perturbativity requirement, for
values of the stop mass splitting smaller than 200 GeV (300 GeV)
for positive (negative) values of the mixing angle.
Only for larger values
of the mass splitting in the stop sector 
viablefrared fixed-point solution of the top-quark mass be
accessible. As shown in Fig.~2, these conclusions depend on the value of
the top quark mass. If the top-quark mass were closer to 180 GeV,
the bounds imposed by  
the Higgs mass limits would become weaker, enlarging the
parameter space consistent with the infrared fixed-point solution.
If, instead, the top-quark mass were closer to 170 GeV, the infrared
fixed point solution would be even more constrained.

To discuss how natural are the large values of the mixing parameters 
needed to reach the infrared fixed point solution,
consider the case where supersymmetry breaking is transmitted
to the observable sector at scales of order $M_{GUT}$.
In this case, the infrared fixed-point solution of the
top-quark mass implies also an infrared fixed point in the
parameters 
$A_t$ and 
${\overline m}^2\equiv m_Q^2+m_U^2+m_{H_2}^2$~\cite{CAOLPOWA,CAWA,ESHA}
\begin{equation}
A_t\approx-1.5 M_{1/2}
\;\;\;\;\;\;\;\;\;\;\;\;\;\;\;\;\;\;\;\;
{\overline m}^2\approx6M_{1/2}^2,
\end{equation}
at scales of the order of the weak scale,
where we have assumed
a common value $M_{1/2}$ for the gaugino masses at $M_{GUT}$
(useful formulae for the most general case can be found
in \cite{CACHOLPOWA,WA}).
For the Higgs potential mass parameter $m_{H_2}^2$, 
one gets from the renormalization
group evolution
$m_{H_2}^2\simeq -3 M_{1/2}^2 + ....$,
while the left- and right-handed stop 
mass parameters increase with $M_{1/2}$,
$m_Q^2\simeq 5.5M_{1/2}^2 + ...$, $m_U^2 \simeq 3.5 M_{1/2}^2 + ....$,
where the ellipses denote a dependence on the values 
of the stop
and Higgs mass parameters at the scale $M_{GUT}$. Hence, 
increasing the stop masses and/or
the gaugino masses, tends to produce large negative values of
$m_{H_2}^2\approx-0.5{\cal O}(M^2_{\tilde t_i})$~\footnote{This can   
only be avoided by having very large values of
the scalar mass parameters at the GUT scale, with very specific
correlations between them \cite{CAWA,CACHOLPOWA}.}. Combining 
these solutions
with the condition of electroweak symmetry breaking,
one finds that large values of $\tilde{A}_t$ can be naturally obtained
at the fixed point. In such cases, however, these large  values of
$\tilde{A}_t$ are negative, implying that the values 
of $\tan\beta$ associated
with the infrared fixed point solutions move to lower values compared to
the case of no mixing.

Figure 3 shows the
bounds on $\tan\beta$ that will be obtained in 
the case that no Higgs boson signal is
found at LEP for $\sqrt{s} = 192$ GeV, implying a 
bound $M_h\simgt98$ GeV~\cite{CAZER}~\footnote{As emphasized
above, in the numerical computations we have lowered the bound
by 2 GeV with respect to the considered Higgs mass limit}.
In that case, even
for large values of the stop mass splitting and the mixing 
parameter, the
bound on $\tan\beta$ resulting 
from the Higgs mass constraints will be stronger
than the perturbativity bounds for values of the heavier stop mass smaller
than 700 GeV (1 TeV), for positive (negative) values of the stop mixing
angle. Hence, as was already emphasized in 
different
works, a run of LEP at $\sqrt{s}=192$ GeV will test most of the parameter
space consistent with the infrared fixed-point solution
\cite{IRFIX,CAWA,ESHA} for a top-quark mass $m^{pole}_t \simlt 175$~GeV.
%It is important to observe, however, that for positive values of the
%mixing angle the upper bound on the Higgs mass at the 
%infrared fixed point solution may be increased by a few GeV.

Finally, Fig. 3
also  shows the bounds that will apply after the final run of LEP, at
$\sqrt{s} = 200$ GeV, assuming a potential lower limit on the Higgs 
boson mass of order
108 GeV~\cite{CAZER}.  It is clear that only  moderate or large values of
$\tan\beta$ will be allowed if the Higgs boson is not found at the final
run of LEP. This will provide a strong motivation for $SO(10)$-type
unification models, in which large values of $\tan\beta$ and Higgs masses of
order $110-120$ GeV are naturally predicted \cite{CAOLPOWA2}.
Of course, from Figs.~1 and 3, one can also
infer the  values of the stop mass parameters consistent
with the infrared fixed-point solution of the top-quark mass, in the 
case that the
Higgs boson is found at the future runs of the LEP collider.

Up to now we have been discussing the situation with 
a large $CP$-odd Higgs boson
mass, $M_A\gg M_Z$. Since for the other MSSM parameters fixed, $M_h$ is maximal
for $M_A\simgt250$ GeV, this is the configuration that is  expected
to yield the smallest lower bound on $\tan\beta$.
Indeed, for smaller values of $M_A$ (for a fixed  stop spectrum and fixed
value of $\tan\beta$) both the coupling to the $Z^0$ boson
and the mass of the lighter $CP$-even Higgs particle  decrease.
For $\tan\beta\simlt3$ and 150 GeV$\simlt M_A\simlt250$ GeV, the
decrease of $M_h$ compensates the drop in the $h^0Z^0Z^0$ coupling, so that
the Higgs boson strahlung cross section actually increases, implying a
bound on $\tan\beta$ stronger than that obtained 
for $M_A\simgt250$ GeV~\cite{CAZER}.
For  values
of $M_A\simlt130$ GeV and values of $\tan\beta\simgt4$, however, this ceases
to be true. In this regime, the $h^0A^0$ associated production cross section
rapidly increases, and this becomes the most efficient channel for
supersymmetric Higgs boson detection. 
%Values of $M_A\simlt80$ GeV are already
%excluded for such values of $\tan\beta$ \cite{WHO?}.
Hence, a small window
for $M_A\simeq {\cal O}$(100~GeV) may still exist, for
which the lower bound on $\tan\beta$  for a given stop 
spectrum may be lower
than the ones presented in this
work (see also \cite{OPAL})~\footnote{This depends, 
however, on the other details of the MSSM
spectrum, since, in this case, constraints from $b\rightarrow\gamma s$ and
$Z^0\rightarrow\bar bb$ come into play.}.
Since this window tends to occur for relatively large values of $\tan\beta$,
for which precise determination of the bounds would require exploration
of the full Higgs boson discovery potential at LEP (and combination of
the results of the four LEP collaborations at the next runs of LEP), we
shall not explore this possibility within this work.

%Another aspect that  is important to emphasize is the one associated with
%the possible appearence of phases in the supersymmetric mass parameters
%$A_t$ and $\mu$. For large values of $M_A$, the Higgs mass depends on the
%modulus of the stop mixing mass parameter, $|A_t-\mu^*/\tan\beta|$. All the
%formulae given above are still valid in the case of complex parameters, by
%making the simple replacement of $\tilde{A}_t$ by $|\tilde{A}_t|$. For fixed
%values of the moduli of the parameters $A_t$ and $\mu$, the Higgs mass
%bound will be of course modified by the presence of phases. However, the
%upper bound of the Higgs mass, and its dependence on the modulus of the
%effective stop mixing mass parameter will remain as in the case of
%real parameters $A_t$ and $\mu$.

It is also instructive to study the impact of precision electroweak
measurements
on the stop mass limits derived above. In Fig. 4, we compare for
$\tan\beta=1.5$  and
$m_t=175$ GeV (close to the infrared fixed point), 
and three
different values of the lighter stop mass $M_{\tilde t_1}$, the regions in the
$(M_{\tilde t_2}, ~\theta_{\tilde t})$ plane allowed by the present limit
on the Higgs boson mass\footnote{As explained before, we take $M_h>88$ GeV
as a conservative estimate in our numerical computation.} (solid lines) and
by precision measurements (shadowed area). To be conservative the precision
data constraints are taken into account by requiring
$\Delta^{stops}\rho\simlt6\times10^{-4}$)~\cite{ERPI}.
The precision measurement bounds are  clear: for $\theta_{\tilde t}=0$
(corresponding to purely right-handed lighter stop), the mass of the heavier
(left-handed) stop is bounded from below
%\footnote{The limit coming from
%the full fit to all precision data \cite{PRECDATA} would depend on
%$M_{\tilde t_1}$ through the dependence of $M_h$ on $M_{\tilde t_1}$.
%We neglect here this small effect.}
(coming from the imposed bound on $\Delta\rho$), but no upper
bound can be set.
For values of $\theta_{\tilde t}\simeq\pi/2$, the constraint on $\Delta \rho$
can be satisfied only by tuning the value of
$\tilde{A}_t$ to be large and of the order of the right-handed
stop mass~\cite{CHCAWA,CH}. For
$\sin2\theta_{\tilde t}\simeq1$, precision measurements put an upper bound on
the heavier stop mass, which, for sufficiently large splitting of the stop
masses, can be lower than the lower bound obtained from the 
limits on the Higgs boson mass.
Hence, for a given value of the heavier stop mass
a non-trivial bound on the lighter stop mass 
can be obtained. In particular, we see from
Fig. 4 that for values of $\tan\beta$ close to the fixed point and values
of the mixing that maximize the Higgs-boson mass, precision measurements
disfavours values of the lighter stop mass below 150 GeV. Notice, however,
that acceptable values of the parameter $\Delta \rho$ can be
obtained by increasing 
the heavier (or the lighter) stop mass. In particular,
close to the fixed point, and for values of the lighter stop mass above 150
GeV, the bounds on the stop  parameters imposed by the limits on
the Higgs boson mass become stronger than the ones coming from precision
data (the opposite is true for large values of $\tan\beta$~\cite{CH}).

A striking result that appears from Fig. 4
is the existence of an effective upper bound on the heavier stop mass
(for fixed $\theta_{\tilde t}$) from the present 
limit on the Higgs boson mass.
It is interesting to understand the situation for values of
$\sin2\theta_{\tilde t}\simeq1$, for which the largest
values of  the Higgs boson
mass are obtained. In this case, the mixing parameter is approximately given by
\begin{equation}
\tilde{A}_t\simeq{M^2_{\tilde t_2}-M^2_{\tilde t_1}\over2m_t}.
\end{equation}
This means that the ratio
${\tilde A}_t/M_{\rm SUSY}\equiv{\tilde A}_t/M_{\tilde t_2}$ grows as
$M_{\tilde t_2}/m_t$ for growing $M_{\tilde t_2}$, leading, by 
Eqs.~(\ref{mhsm}),~(\ref{escala}), to negative contributions to $M_h$, which
rapidly overcome the positive logarithmic dependence on $M_{\tilde t_2}$.
For very large stop mass splitting, of course,
Eq.~(\ref{mhsm}) is not applicable, and the exact value should be obtained
by using the whole renormalization group improved effective potential
\cite{CAESQUWA}.

As emphasized above, the perturbativity bounds depend also on the physics
at scales larger than the supersymmetric particle masses and, hence, are
model-dependent.
Adding new matter multiplets with non-trivial 
$SU(3)\times SU(2)\times U(1)$ quantum numbers at some
intermediate scale $M_I$,
e.g. extra ${\bf 5}+{\bf \bar5}$
and/or ${\bf 10}+{\bf \overline{10}}$ matter representations (having no
interactions
with the ordinary matter in the superpotential),  
decreases the lower perturbativity limit on
$\tan\beta$. 
This is easy to understand, by noting that above the scale $M_I$,
$\alpha_s$ becomes less asymptotically free (i.e. goes up steeper) and,
therefore, has a stronger damping effect on 
the top quark Yukawa coupling, thus
allowing for a larger  initial value at $Q=m_t$ (and, hence,
lower $\tan\beta$).
Of course, adding more extra representations at lower scale $M_I$ allows for
smaller $\tan\beta$. One could hope, therefore, that with a suitable number
of extra representations at some scale $M_I$ one can reach a lower limit on
$\tan\beta$ smaller than 1 and at the same time satisfy also the Higgs mass
limit (see Fig. 1). However, for a given scale $M_I$ the number of extra
representations is limited by perturbativity of the gauge couplings.
With all one-loop threshold corrections to the relation (\ref{eqn:mtyukrel})
we find that for $M_I\simgt10^5$ GeV one can afford
at most five ${\bf 5}+{\bf \bar5}$ representations (or two
${\bf 5}+{\bf \bar5}$ and one ${\bf 10}+{\bf \overline{10}}$ representations)
for
relatively heavy ($\simgt1.5$ TeV) sparticle spectra. For sparticle spectra
$\simlt1$ TeV only four ${\bf 5}+{\bf \bar5}$ representations (or one
${\bf 5}+{\bf \bar5}$ and one ${\bf 10}+{\bf \overline{10}}$  representations)
are allowed. As a result, the perturbativity  limit on $\tan\beta$
can approach 1  only for very heavy sparticle spectra
($\simgt2$ TeV) and/or large mixing stop mass parameters.
This is illustrated in Fig.~5.

In  very interesting works~\cite{ELOL,ABAL}, it has been
shown that, in the minimal supergravity model, the infrared fixed-point
solution is already ruled out by the requirement of having a phenomenologically
acceptable amount of dark matter
and/or avoiding charge- or colour-breaking minima.
It is important to notice, however, that in
general supergravity models, the masses of sfermions with different quantum
numbers may be different. In particular, there might be no correlation between
the slepton, Higgs, neutralino and squark masses.
Without these correlations, it is
difficult to relate the neutralino annihilation cross section to the Higgs and
stop spectrum and hence, the fixed-point solution cannot be ruled out by these
considerations. Moreover, even if the correlations between sparticle
masses were  similar to the ones present in the minimal supergravity
model, a tiny violation of $R$-parity would be sufficient to suppress these
cosmological constraints on the infrared fixed-point solution
and to avoid dangerous colour- or charge-breaking minima~\cite{ABAL}.

In another independent work, it has been noted \cite{CEP} that the amount
of fine tuning~\cite{BAGI}
increases for low values of $\tan\beta$ close to the fixed-point. 
This is specially the case
for the large values of the stop masses
that are necessary to approach 
the fixed-point solution. If a Higgs particle is found
in the next runs of the LEP collider, it would be interesting to investigate
the conditions necessary to obtain a spectrum consistent with the 
fixed-point solution in a natural way. 
If it is not found, the fixed-point solution
will be ruled out by solid experimental data.

Another cosmologically interesting scenario, which demands Higgs masses
in the range of LEP2, is electroweak baryogenesis \cite{CQW}. The
realization of this scenario, however, demands a light stop and
relatively small values of the stop mixing. As 
follows from the present analysis of the constraints 
on the stop sector imposed by the
Higgs boson mass limits (and precision data),
the above requirements can only be satisfied either for moderate
values of $\tan\beta$, or for very large values of the heavier
stop mass.
In fact, for values of the heavier stop mass at most of the order 
of~2~TeV, a lower bound on $\tan\beta \simgt 2$ can already
be obtained in this particular case. Hence, this scenario is not
consistent with the infrared fixed-point solution.

Let us finish this discussion by mentioning that in this work we have
assumed the absence of any extra Higgs-like states in the low-energy spectrum.
For instance, the presence of a singlet~\cite{EXTRASINGL},
with a tree-level
superpotential coupling $\lambda S H_1 H_2$ would induce a tree-level
quartic coupling for the lighter 
CP-even Higgs boson proportional to $\lambda^2\sin^22\beta$
\cite{ELETAL}. This tree-level contribution would become 
most important for low
values of $\tan\beta$ and could only 
be constrained by perturbativity limits on the
coupling $\lambda$. If such a singlet were present in the low-energy spectrum,
the bounds on $\tan\beta$ and on
the stop mass parameters would be considerably modified.

\vskip 0.3cm
~\\
{\bf Acknowledgements} The work of P. Ch. and S. P. has been partly supported
by the Polish State Committee for Scientific Research grant 2 P03B 030 14.
M.C. and C.W. would like to thank J.R. Espinosa, S. de Jong, P.~Janot,
H. Haber, P. Igo-Kemenes and G. Weiglein  for useful discussions.

%\newpage

\begin{figure}
\psfig{figure=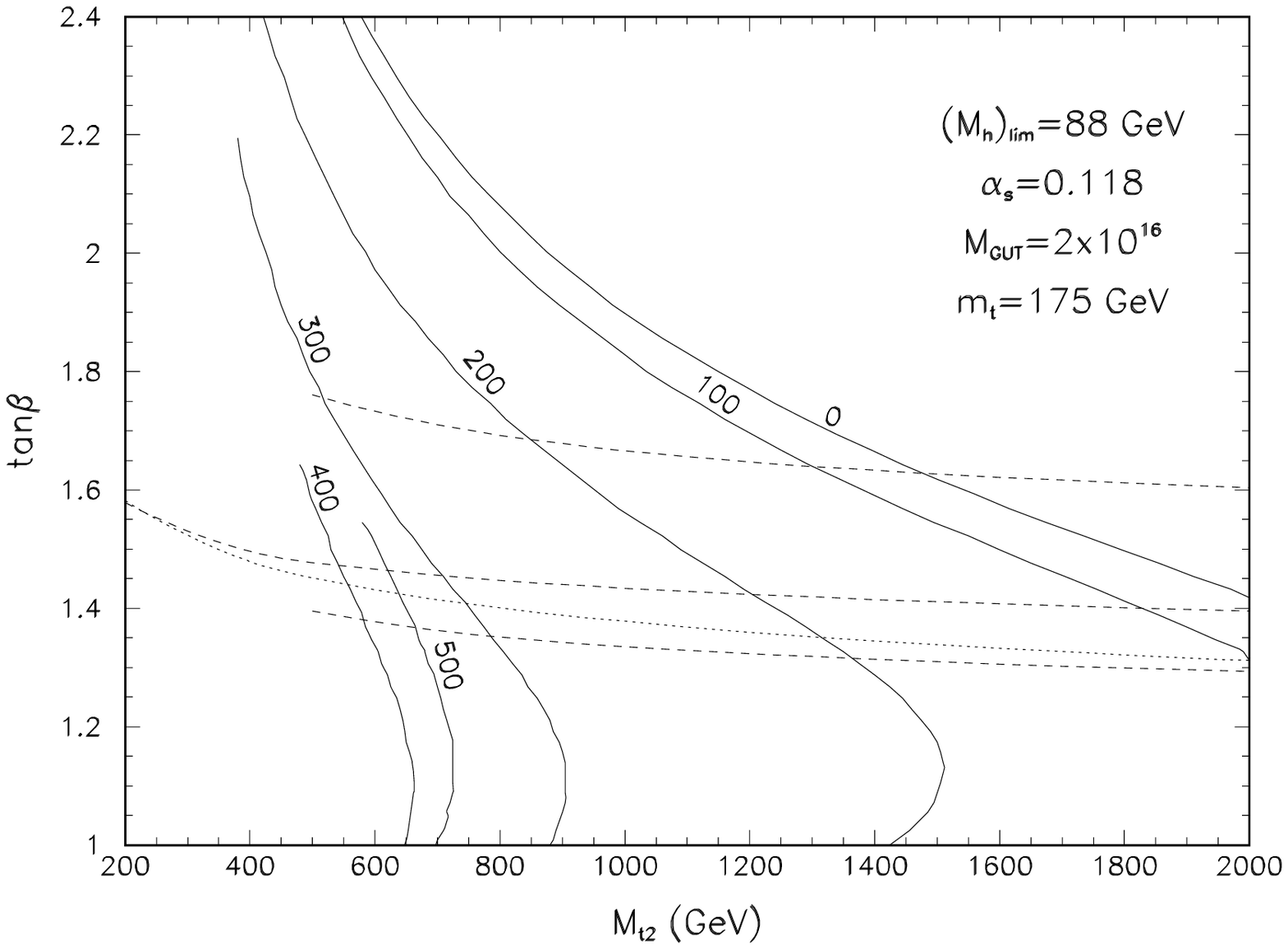,width=15.0cm,height=15.0cm}
\vspace{1.0truecm}
%height=13cm,bbllx=4.5cm,bblly=.cm,bburx=14.cm,bbury=13cm}}
\caption{Bounds on $\tan\beta$ obtained for $m_t^{pole}=175$ GeV and for
a lower bound on the Higgs boson mass $M_h>88$ GeV,
as a function of the heavier stop
mass, for different values of the stop mass splitting 
$\Delta M_{\tilde t} = 0$--500 GeV (solid lines).
Also plotted here are the top Yukawa coupling perturbativity
bounds 
for the case of heavy gluino ($m_{\tilde g} = M_{\tilde t_2}$) 
for  $\Delta M_{\tilde t} = 400$ GeV
and 
$\sin2\theta_{\tilde t}=\pm1$ (upper-lower dashed lines), 
and for heavy
gluino (center-dashed lines) and light gluino ($m_{\tilde g} = 200$ GeV)
(dotted line)  for $\Delta M_{\tilde t} = 0$.}
\end{figure}

\begin{figure}
\psfig{figure=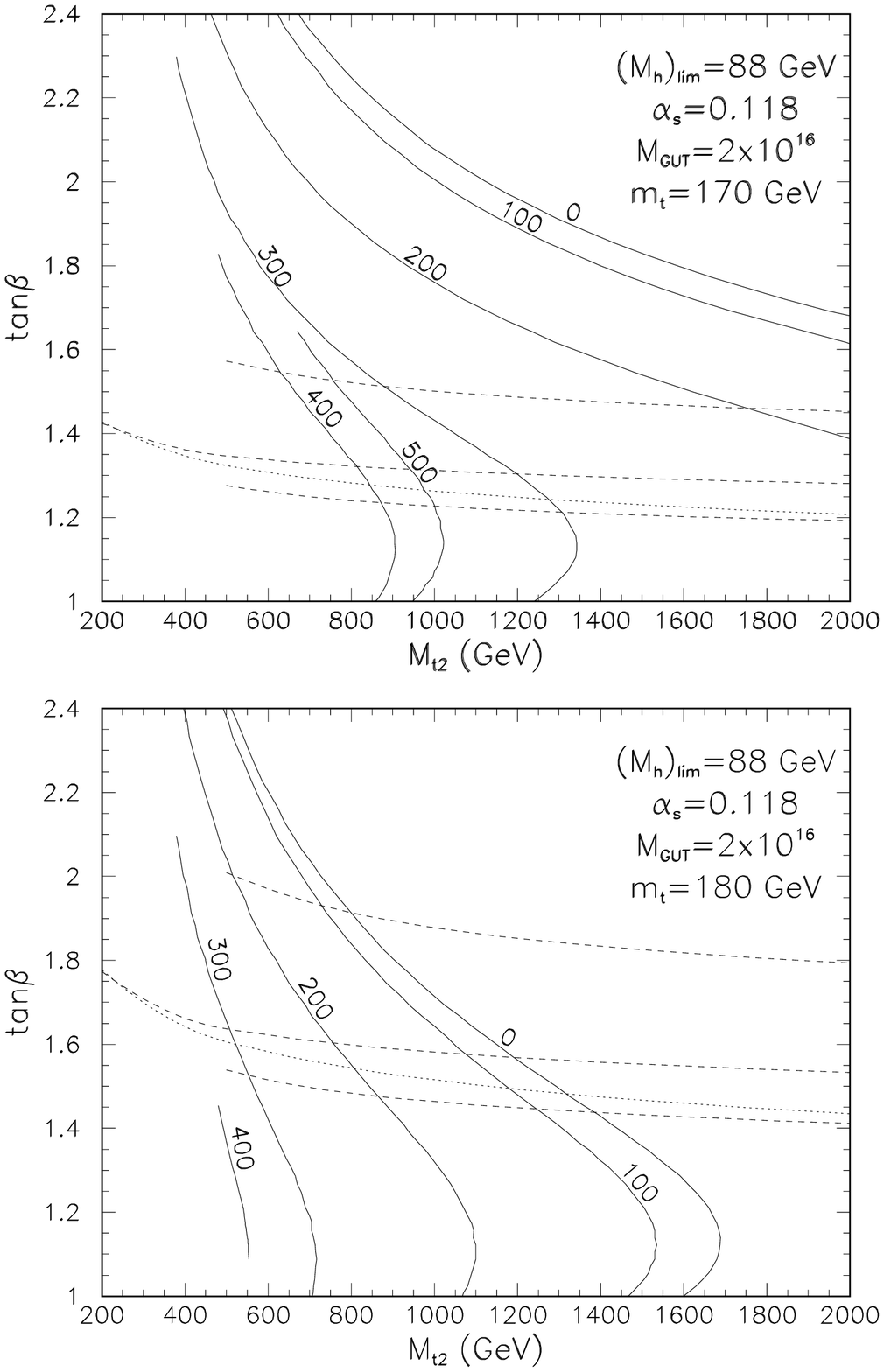,width=12.0cm,height=19.0cm}
\vspace{1.0truecm}
%height=13cm,bbllx=4.5cm,bblly=.cm,bburx=14.cm,bbury=13cm}}
\caption{The same as Fig. 1, but for $m_t^{pole} = 170$ GeV
and $m_t^{pole}= 180$ GeV.}
\end{figure}

\begin{figure}
\psfig{figure=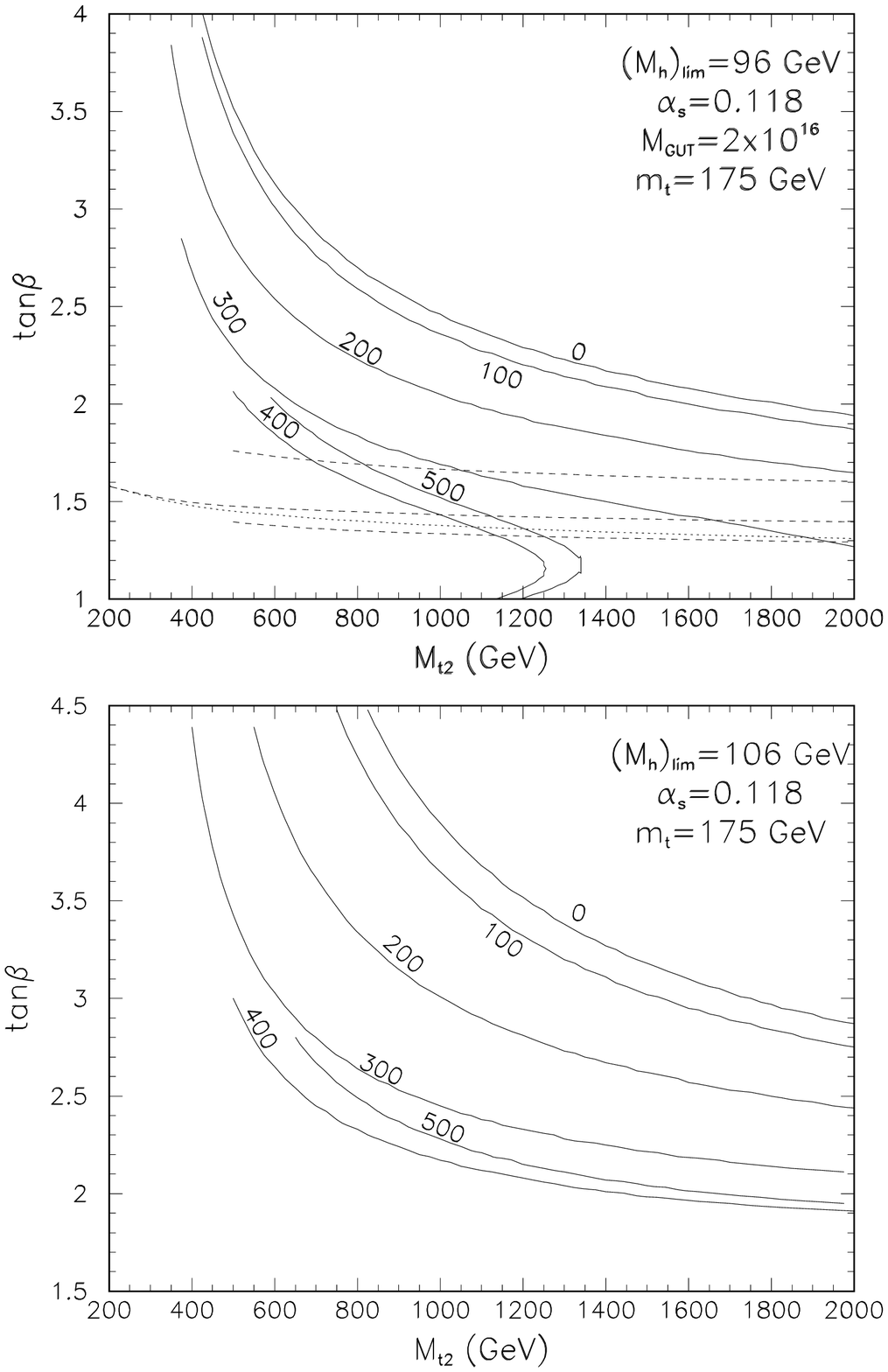,width=12.0cm,height=19.0cm}
\vspace{1.0truecm}
%height=13cm,bbllx=4.5cm,bblly=.cm,bburx=14.cm,bbury=13cm}}
\caption{The same as Fig. 1, but for  lower bounds on the
Higgs boson mass of 96 GeV and 106 GeV.}
\end{figure}

\begin{figure}
\psfig{figure=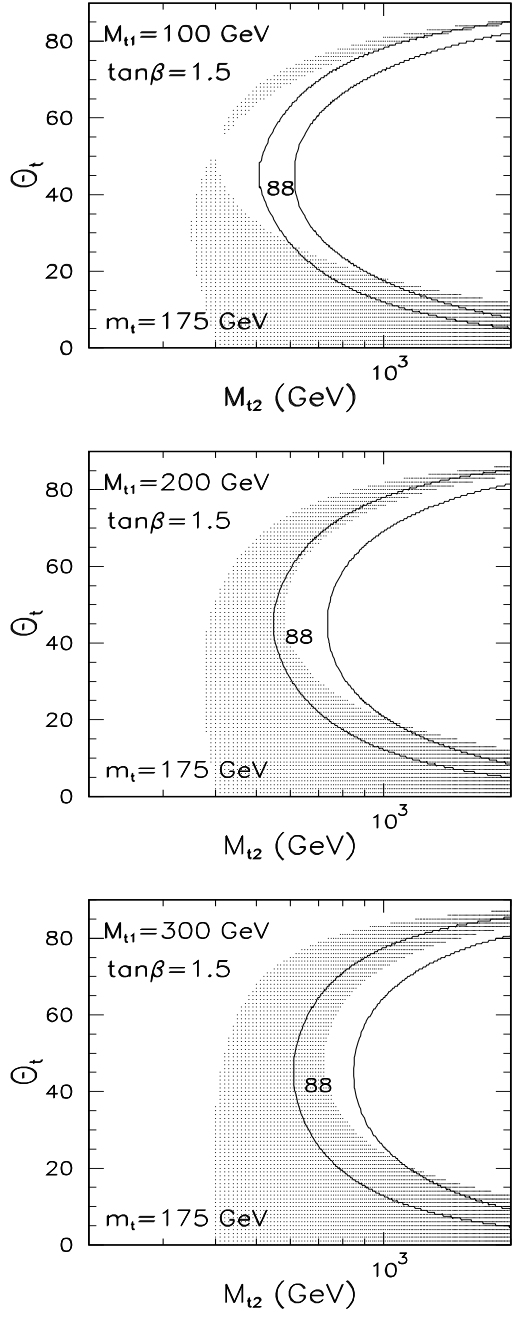,width=7.50cm,height=18.0cm}
\vspace{1.0truecm}
%height=13cm,bbllx=4.5cm,bblly=.cm,bburx=14.cm,bbury=13cm}}
\caption{Bounds coming from the constraints on the parameter
$(\Delta \rho)^{stops}$ (shadowed region) in the heavier stop mass--stop
mixing angle plane, for values of $\tan\beta = 1.5$,
and for different values
of the lighter stop mass $M_{\tilde t_1} = 100$, 200 and 300 GeV.
Also shown here are the bounds obtained from the present
limit on the Higgs boson mass (regions between solid lines).}
\end{figure}

\begin{figure}
\psfig{figure=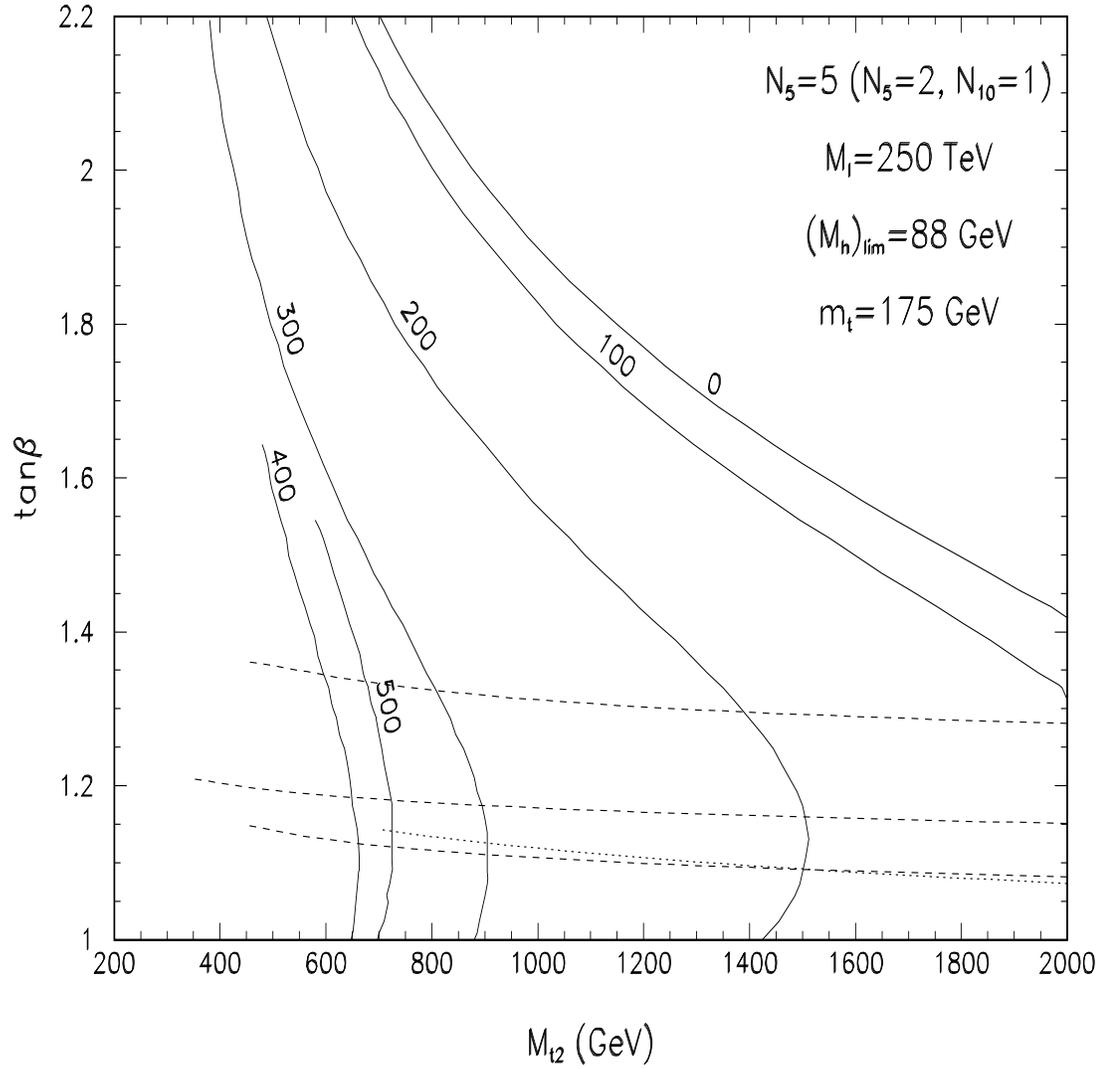,width=15.0cm,height=15.0cm}
\vspace{1.0truecm}
%height=13cm,bbllx=4.5cm,bblly=.cm,bburx=14.cm,bbury=13cm}}
\caption{The same as Fig. 1, but with five
additional ${\bf 5}+{\bf \bar5}$ pairs
(or equivalently, two ${\bf 5}+{\bf \bar5}$  pairs and
one ${\bf 10}+{\bf \overline{10}}$ pair)  added at the
scale of 250 TeV.}
\end{figure}

\end{document}